\documentclass[preprint,showkeys,preprintnumbers,amsmath,amssymb,aip,apl]{revtex4}

\usepackage{graphicx}
\usepackage{dcolumn}
\usepackage{bm}

\begin{document}


\title{Modal ``self-coupling" as a sensitive probe for nanomechanical detection}

\author{M. Defoort}\affiliation{Institut N\'eel, CNRS et Universit\'e Joseph Fourier, BP 166, 38042 Grenoble Cedex 9, France}
\author{K.J. Lulla}\affiliation{Institut N\'eel, CNRS et Universit\'e Joseph Fourier, BP 166, 38042 Grenoble Cedex 9, France}
\author{C. Blanc}\affiliation{Institut N\'eel, CNRS et Universit\'e Joseph Fourier, BP 166, 38042 Grenoble Cedex 9, France}
\author{O. Bourgeois}\affiliation{Institut N\'eel, CNRS et Universit\'e Joseph Fourier, BP 166, 38042 Grenoble Cedex 9, France}
\author{A. D. Armour}\affiliation{School of Physics and Astronomy, University of Nottingham, Nottingham NG7 2RD, United Kingdom}
\author{E. Collin}\affiliation{Institut N\'eel, CNRS et Universit\'e Joseph Fourier, BP 166, 38042 Grenoble Cedex 9, France}

\email{eddy.collin@grenoble.cnrs.fr}

\date{\today}

\begin{abstract}
We present a high-sensitivity measurement technique for mechanical nanoresonators.
Due to intrinsic nonlinear effects, different flexural modes of a nanobeam can be coupled while driving each of them on resonance. This mode-coupling scheme is dispersive and one mode resonance shifts with respect to the motional amplitude of the other. The same idea can be implemented on a {\it single} mode, exciting it with two slightly detuned signals.
This two-tone scheme is used here to measure the resonance lineshape of one mode through a frequency shift in the response of the device.
The method acts as an amplitude-to-frequency transduction which ultimately suffers only from phase noise of the local oscillator used and of the nanomechanical device itself.
We also present a theory which reproduces the data without free parameters.
\end{abstract}

\maketitle

Nanoelectromechanical structures (NEMS) have been studied intensively both because of their potential applications and their suitability for investigating different areas of fundamental physics.  Over recent years NEMS have shown great promise as ultrasensitive force sensors, they have been used to detect ever smaller masses \cite{RoukesNatNanoT}, as well as being used to measure individual spins \cite{RugarIBM}. NEMS also possess relatively strong nonlinearities which means that they are very suitable for investigating fundamental aspects of nonlinear dynamics \cite{aldridge,Lifshitz}. Moreover, it has also been possible to access the quantum regime for these systems \cite{ClelandNat,Poot}.

Detecting the tiny motion of nanomechanical devices is usually a challenging experimental issue, especially for high (radio-)frequency modes.
An interesting approach involving {\it two modes} of the same structure was presented in Refs. \cite{nlincoupl}: due to (geometrical) nonlinearities, the motional amplitude of the mode under study shifts the resonance position of another mode, used as detector. This ``mode-coupling" scheme has the potential to be extremely efficient, since frequencies are relatively easy to measure with high accuracy.

Other kinds of multiple-tone driving schemes have been developed over the years, tackling various issues of signal processing: with e.g. the parametric frequency tuning of a mode \cite{paramtune}, the frequency comb generation \cite{yamaguchi}, or the audio-frequency mechanical mixing \cite{mix}.
In particular, two-tone driving of a {\it single mode} of a mechanical device has already been used for amplifying purposes \cite{Buks}. Here, we propose to adapt the mode-coupling scheme to a single mode with a dedicated two-tone detection method (with a probe and a drive signal).
This ``self-coupling" method essentially converts the amplitude of the motion into a frequency. Provided one can measure the resonance position of the response to the probe signal with high accuracy (i.e. using a device with a high quality factor and a local oscillator with very good frequency stability), the method should only be limited by the phase noise of the setup.

\begin{figure}[t!]
\includegraphics[height=7.0 cm]{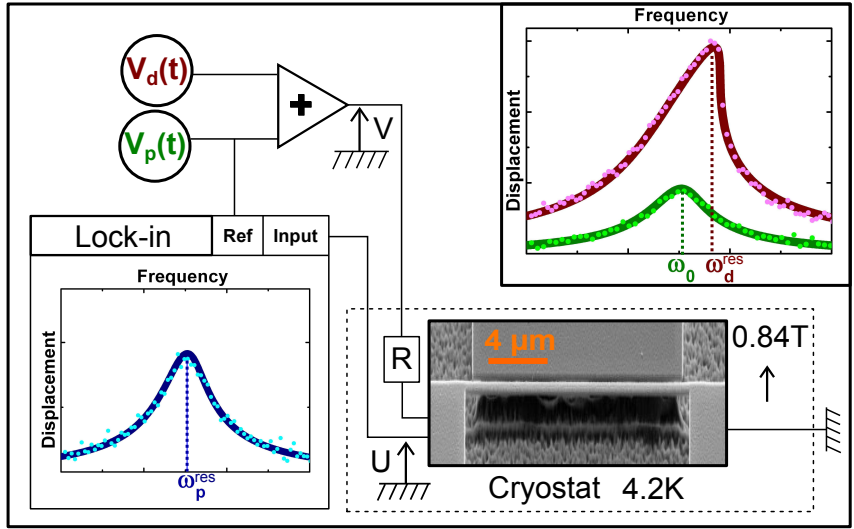}
\caption{\label{Setup} (Color Online) Schematic of the experimental setup. A SEM picture of the sample is shown, with an unused gate at the top.
Two voltage signals are added and converted to a current $I(t)$ through a $R=1~$k$\Omega$ bias resistance. With the field $B=0.84~$T, it generates a superposed force $F(t)=F_d(t)+F_p(t)$, with drive and probe components. The response to the probe is measured with a lock-in amplifier (detection inset, bottom left). The upper right inset shows examples of the measured response of the device separately to the drive (upper points) and to the probe (lower points), together with fits (full lines). Positions of the maximum of each response are indicated by dashed lines with the corresponding frequencies $\omega_p^{res}$, $\omega_0$ and $\omega_d^{res}$.
}
\end{figure}

The experiments are performed on a 15$~\mu$m long silicon-nitride doubly-clamped nanobeam (Fig. \ref{Setup}). This material has been demonstrated to be an excellent choice for nanomechanical purposes, leading to high frequency and high quality factor ($Q$) devices \cite{Parpia}. The structure is 250$~$nm x 100$~$nm wide and thick, covered with 30$~$nm of aluminum. The fabrication techniques are described in Ref. \cite{JLTP}. The sample is placed in  cryogenic vacuum at 4.2$~$K.

We use a magnetomotive scheme to drive and detect the motion \cite{Sensor,RSI}. The device is excited with a Laplace force $F(t)=\xi I(t) l B$, where $l$ is the length of the beam, $\xi$ a mode-dependent number and $B$ the magnetic field. The current is given by $I(t)=V(t)/(R+r)$ where $V(t)$ is the drive voltage applied, $r$ is the electrical resistance of the structure (about 100$~\Omega$) and $R$ the bias resistance.
The out-of-plane motion is detected through the voltage $U(t)$ induced by the magnetic flux cut by the beam's distortion.

We excite the first out-of-plane flexural mode of the beam with two slightly detuned signals added together (Fig. \ref{Setup}).
The resulting force is $F(t)=F_d(t)+F_p(t)$, with $F_d(t)=F_{d,0}\cos(\omega_d t+\phi_d)$ and $F_p(t)=F_{p,0}\cos(\omega_p t+\phi_p)$ the
drive and probe components respectively. The induced voltage $U(t)$ is measured with a lock-in detector referenced on the probe signal $F_p(t)$ at $\omega_p$, leading thus to the probe response only.

The nanobeam's mechanics is nonlinear due to the axial stress induced by the elongation under motion \cite{Lifshitz,Nayfeh,nlincoupl,Kunal}.
The motion of the first flexural mode under a force $F(t)$ is described by the well-known Duffing equation:
\begin{eqnarray}
&& \ddot{x} + \Delta\omega \, \dot{x} + \omega_0^2\,x(1 + \kappa\,x^2)= \frac{F(t)}{m}, \label{eq:eom}
\end{eqnarray}
with $\frac{\omega_0}{2\pi}=6.98~$MHz the resonance frequency of the mode, $\frac{\Delta\omega}{2\pi}=550~$Hz its linewidth, {\it m} the mass and $\kappa$ the non-linear Duffing parameter. The coordinate $x(t)$ characterizing the amplitude of the motion is defined here as the average displacement along the mode shape, such that the mode mass equals the bare mass of the beam.

In the absence of a second drive tone ($F_{p,0}=0$), Eq.\ (\ref{eq:eom}) leads to the standard Duffing response. Written in terms of $A_d$, the complex amplitude of the motion at frequency $\omega_d$, this takes the form
\begin{eqnarray}
\label{Duff}
&& A_d\,=\,\frac{F_{d,0}\,e^{i\,\phi_d}}{2\,m\,\omega_0}\times \frac{1}{\beta\,\left| A_d \right|^2\,-\,\delta\omega\,+\,i\,\frac{\Delta\omega}{2}},
\end{eqnarray}
where $\beta =\frac{3}{8}\omega_0\,\kappa$ is the frequency pulling factor and $\delta\omega=\omega_d-\omega_0$ is the detuned angular frequency (in Radian/s) between the drive and the natural (linear) frequency of the beam.
In the limit of very weak drives, $A_d$ is small and Eq.\ (\ref{Duff}) reduces to the standard Lorentzian expression. For larger values of the drive, the system displays a characteristic frequency pulling and the response is peaked with respect to the frequency $\omega_d$ at a value $\omega_d^{res}= \omega_0 + \beta \left|A_d\right|^2$ (see Fig. \ref{Setup} top right inset).
However, for a sufficiently strong drive the amplitude reaches a threshold value $A_c$ and the solution becomes bistable \cite{Landau,Nayfeh}.

\begin{figure}[t!]
\includegraphics[height=7.0 cm]{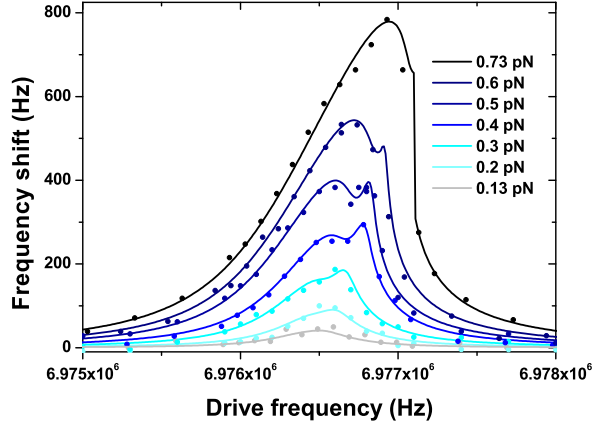}
\caption{\label{CouplShift} (Color online) Frequency shift $(\omega_p^{res}-\omega_0)/(2 \pi)$ of the first mode resonance $A_p$ excited at $\omega_p$ as a function of drive frequency $\omega_d$, for different amplitudes of the drive $F_{d,0}$. The probe force component $F_{p,0}$ was 0.2$~$pN (arising from a current of 13$~$nA), generating itself an amplitude of 1$~$nm. Lines are the theoretical predictions; the scattering of points along the vertical axis is due to the finite resolution of the experimentally determined resonance position (from our fits). }
\end{figure}

Applying two frequency-detuned drives around $\omega_0$ leads to new oscillating solutions for the system \cite{Buks}. We define the detuning between the probe and drive frequency as $\delta\,=\,\omega_p\,-\,\omega_d$ which we assume to be small  $\delta\,\ll\,\omega_d\,,\,\omega_p$ and we also assume that both frequencies are close to resonance $\delta \omega\,\ll\,\omega_d\,,\,\omega_p$.
The Duffing non-linearity in the beam means that in general the amplitude of the motion at the probe frequency, $A_p$, and that at the drive frequency, $A_d$, are not independent of each other.

In the present paper, we consider the case where the probe excitation is weak so that $A_p$ always remains small whilst the drive excitation can be relatively strong so that the amplitude $A_d$ ranges from the Lorentzian regime up to the onset of bistability. Assuming a weak probe signal, we can work to linear order in the response at the probe frequency, $A_p$, so that the response at $\omega_d$ is independent of $A_p$, and follows Eq.\ (\ref{Duff}).
However, as the drive excitation can be relatively large the probe response at $\omega_p$ depends on $A_d$ in a non-trivial way,
\begin{multline}
\label{Ar}
A_p\,=\,\frac{F_{p,0}\,e^{i\,\phi_p}}{2\,m\,\omega_0}\times
\frac{1}{2\,\beta\,\left|A_d\right|^2\,-\,\delta\omega\,-\,\delta\,+\,i\,\frac{\Delta\omega}{2}\,-\frac{\beta^2\,\left|A_d\right|^4}{2\,\beta\,\left|A_d\right|^2\,-\,\delta\omega\,+\,\delta\,-\,i\,\frac{\Delta\omega}{2}}}.
\end{multline}
Our assumption that the amplitude of the response at the pump frequency is small means that unlike Eq.\ (\ref{Duff}), there is no dependence on the amplitude $A_p$ in the right hand side of Eq.\ (\ref{Ar}).

Experimentally, we sweep $\omega_d$ for a given drive force $F_{d,0}$ and measure the position of the resonance peak $A_p$ at $\omega_p=\omega_p^{res}$ (Fig. \ref{CouplShift}). The frequency position of this peak $\omega_p^{res}$ can be computed numerically from Eq.\ (\ref{Ar}), leading to the full lines in Fig. \ref{CouplShift}. Note that all mechanical parameters of the device were carefully measured using the calibration method described in Ref. \cite{RSI}, so the calculation has no free parameters. We thus demonstrate that the resonance lineshapes obtained from Eq.\ (\ref{Ar}) fit perfectly the data even for the regime where the response at the drive frequency is nonlinear. Our resolution here is only limited by our simple fitting procedure for the resonance position of the probe, which is typically within $\pm 20~$Hz.

\begin{figure}[t!]
\includegraphics[height=7.0 cm]{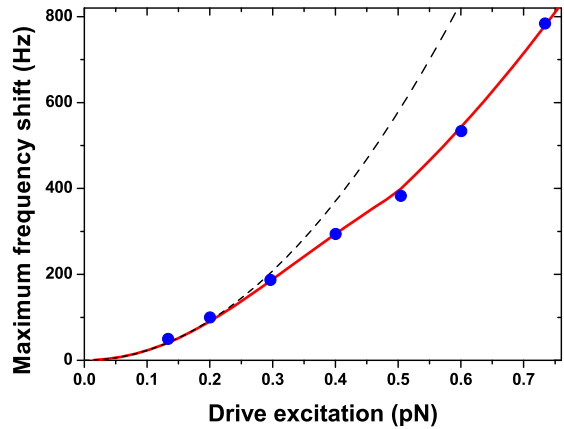}
\caption{\label{MaxShift} (Color online) Maximum heights of the peaks measured in Fig. \ref{CouplShift} (in terms of $A_p$ frequency shifts) as a function of drive force $F_{d,0}$. A numerical calculation using Eq.\ (\ref{Ar}) is shown with the full (red) line. A comparison with the analytic first order behavior, from Eq.\ (\ref{Omr}), is also shown (dashed line).
}
\end{figure}

From Fig. \ref{CouplShift} we can extract the maximum height of each curve, in terms of frequency shifts, with respect to $F_{d,0}$.
This is shown in Fig. \ref{MaxShift} together with a numerical calculation using Eq.\ (\ref{Ar}) (full line).
A particularly simple result is obtained in the limit of small $F_{d,0}$, where the $\beta^2\,\left|A_d\right|^4$ term in Eq.\ (\ref{Ar}) can be neglected. In this limit the amplitude $A_p$ is thus the standard  Lorentzian (of a linear oscillator), centered at the frequency:
\begin{eqnarray}
\label{Omr}
&& \omega_p^{res}=\omega_0 + 2\beta \left|A_d\right|^2 .
\end{eqnarray}
Note the factor of 2 compared to $\omega_d^{res}$. Inserting the maximal amplitude of the $A_d$ peak resonance, Eq.\ (\ref{Duff}), into this expression leads to the parabolic curve in Fig. \ref{MaxShift} (dashed line), which matches the full numerical result (given as a full line in Fig. \ref{MaxShift}) for small drives. Expression Eq.\ (\ref{Omr}) is similar to what is obtained for the mode coupling scheme  (where for weak excitations the frequency shift in one mode depends quadratically on the amplitude of a second mode \cite{nlincoupl,Kunal,vanderZant}), though here only a single mode is involved.

 Turning now to the application of this method for the detection of small amplitudes of mechanical motion, the figure of merit is essentially $2 \beta$, the frequency shift per unit amplitude squared. For our silicon nitride doubly-clamped beam we obtain a value of about $100~$Hz/nm$^2$ which is 15 times larger than the inter-mode coupling value reported in Ref. \cite{Kunal}, for a similar device. In other words, a motion of 1$~$nm which induces a voltage of only 450$~$nV at 7$~$MHz in the standard magnetomotive scheme, shifts the resonance peak of the probe signal by 20\% of its linewidth.

In conclusion, we presented a method enabling the accurate measurement of a nanomechanical mode resonance peak.
The experiment was performed using a doubly-clamped nanobeam resonating at 7$~$MHz, and the complete theory describing the results is given.
The technique uses two signals slightly frequency-detuned, mimicking the mode coupling scheme though instead using only a single mode. This two-tone ``self-coupling" detection converts the motion into a frequency through the measurement of the resonance position of the linear response to one of the tones. It is thus ultimately only limited by the phase noise of the whole setup. We believe this scheme could be of interest to many areas of applied physics, even outside of the nanomechanics field.

The authors want to thank H. Godfrin for discussion, and J. Minet and C. Guttin for help in setting up the experiment. We also thank T. Crozes and T. Fournier for help in the fabrication of samples. We acknowledge the support from MICROKELVIN, the EU FRP7 low temperature infrastructure grant 228464 and of the 2010 ANR French grant QNM no 0404 01.

\end{document}